\documentclass[letters,usenatbib]{mnras}

\usepackage[T1]{fontenc}

\DeclareRobustCommand{\VAN}[3]{#2}
\let\VANthebibliography\thebibliography
\def\thebibliography{\DeclareRobustCommand{\VAN}[3]{##3}\VANthebibliography}

\usepackage{graphicx}	
\usepackage{amsmath}	
\usepackage{amssymb}	
\usepackage{newtxtext,newtxmath}

\title[King Ghidorah Supercluster]{King Ghidorah Supercluster: Mapping the light and dark matter in a new supercluster at z = 0.55 using the Subaru Hyper Suprime-Cam}

\author[R. Shimakawa]{
Rhythm Shimakawa,$^{1}$\thanks{E-mail: rhythm.shimakawa@nao.ac.jp}\thanks{NAOJ Fellow}
Nobuhiro Okabe,$^{2}$
Masato Shirasaki$^{1,3}$
and Masayuki Tanaka$^{1}$
\\
$^1$National Astronomical Observatory of Japan (NAOJ), National Institutes of Natural Sciences, Osawa, Mitaka, Tokyo 181-8588, Japan\\
$^2$Graduate School of Advanced Science and Engineering, Hiroshima University, 1-3-1 Kagamiyama, Higashi-Hiroshima, Hiroshima 739-8526, Japan\\
$^3$The Institute of Statistical Mathematics, Tachikawa, Tokyo 190-8562, Japan
}

\date{Accepted 2022 November 21. Received 2022 November 2; in original form 2022 August 31}

\pubyear{2022}

\begin{document}
\label{firstpage}
\pagerange{\pageref{firstpage}--\pageref{lastpage}}
\maketitle

\begin{abstract}
This paper reports our discovery of the most massive supercluster, termed the King Ghidorah Supercluster (KGSc), at $z=0.50-0.64$ in the Third Public Data Release of the Hyper Suprime-Cam Subaru Strategic Program (HSC-SSP PDR3) over 690 deg$^2$, as well as an initial result for a galaxy and dark matter mapping.
The primary structure of the KGSc comprises triple broad weak-lensing (WL) peaks over 70 comoving Mpc.
Such extensive WL detection at $z>0.5$ can only currently be achieved using the wide-field high-quality images produced by the HSC-SSP.
The structure is also contiguous with multiple large-scale structures across a $\sim400$ comoving Mpc scale. 
The entire field has a notable overdensity ($\delta=14.7\pm4.5$) of red-sequence clusters.
Additionally, large-scale underdensities can be found in the foreground along the line of sight.
We confirmed the overdensities in stellar mass and dark matter distributions to be tightly coupled and estimated the total mass of the main structure to be $1\times10^{16}$ solar masses, according to the mock data analyses based on large-volume cosmological simulations.
Further, upcoming wide-field multi-object spectrographs such as the Subaru Prime Focus Spectrograph may aid in providing additional insights into distant superclusters beyond the 100 Mpc scale.
\end{abstract}





\section{Introduction}
\label{s1}

Superclusters are known to be the largest coherent structures in the universe. 
They contain numerous galaxy clusters with sizes greater than $\sim100$ $h^{-1}$Mpc, which are deemed to have originated from the initial density perturbations on a large scale \citep{deVaucouleurs1953,Gregory1978,Kirshner1981,Oort1983}.
They may provide useful clues regarding the nature of the density fluctuations in the early universe \citep{Zel'dovich1970,White1979,Einasto2021} and the hierarchical evolution of galaxies and large-scale structures (see, e.g., \citealt{Einasto2014,Galametz2018,Paulino-Afonso2020} for galaxy properties and also \citealt{Busha2003,Nagamine2003,Einasto2019} for large-scale structures).
Hence, investigating the mass distributions of both galaxies and the underlying dark matter in superclusters is crucial to better comprehend the mass assembly history of baryonic matter and dark matter in the universe.

Past spectroscopic campaigns and weak lensing (WL) studies have successfully constrained the masses and dynamical states of superclusters at low redshifts.
These studies have resolved complex matter structures in superclusters to be $\sim10^{15}$--$10^{17}$ M$_\odot$ and have revealed their close correlations with galaxy number densities \citep{Geller1999,Reisenegger2000,Bardelli2000,Heymans2008,O'Mill2015,Higuchi2020}.
They have also discovered morphological diversities among superclusters, further suggesting the varying evolutionary histories \citep{Shandarin2004,Einasto2011,Einasto2014}.
Given the recent development of gigapixel-level prime focus cameras for use in large ground-based telescopes \citep{Miyazaki2018,Ivezic2019}, the next step involves extending such analyses to more distant superclusters.
However, to date, the number of distant supercluster discoveries remains limited \citep{Gunn1986,Lubin2000,Nakata2005,Gal2008,Tanaka2009,Mei2012,Galametz2018,Paulino-Afonso2018}.

In this context, we are currently conducting a systematic supercluster (and void) search at $z>0.3$ based on the Hyper Suprime-Cam Subaru Strategic Program (HSC-SSP; \citealt{Aihara2018}), and this paper reports an initial result focusing on the most massive supercluster found at $z=0.55$ within the samples considered in our previous study \citep[\S\ref{s2}]{Shimakawa2021a}.
This research is particularly motivated by the desire to investigate galaxy and dark matter overdensities and their spatial relations in the supercluster based on WL analysis (\S\ref{s3}).
Combined with cosmological simulations, we also estimate the total mass of the supercluster (\S\ref{s4}).

For this study, we adopt the AB magnitude system \citep{Oke1983} and the \citet{Chabrier2003} initial mass function. 
In line with relevant previous studies \citep{Takahashi2017,Shirasaki2017,Shimakawa2021a,Shimakawa2021b}, we assume the following values for cosmological parameters, $\Omega_M=0.279$, $\Omega_\Lambda=0.721$, and $h=0.7$, in a flat lambda cold dark matter model, which are consistent with those obtained from the WMAP nine-year data \citep{Hinshaw2013}.


\clearpage
\section{Data and target selection}
\label{s2}

This study was based on the data from the Third Public Data Release of the HSC-SSP (HSC-SSP PDR3; \citealt{Aihara2022}).
The data were collected over 278 nights and reduced using the {\tt hscPipe} software (version~8; \citealt{Bosch2018}).
This research focused on the data covering 690 deg$^2$ in the $grizy$ bands at complete depth (e.g., $i>26$ mag at $5\sigma$ limiting magnitudes).
We employed a projected density map to obtain magnitude limited samples of galaxies \citep{Shimakawa2021a}, along with the red-sequence cluster catalogue \citep{Oguri2018}; however, we used updated versions for the HSC-SSP PDR3\footnote{\url{https://hsc-release.mtk.nao.ac.jp/doc}}.
As complete descriptions of these databases are available in their respective papers, this section only presents a summary.

\citet{Shimakawa2021a} obtained the number densities of $i$-band magnitude limited galaxies ($i<23$) in the projected space of the following redshift interval $\Delta z=0.1$ from $z=0.3$ to $z=1$ within circular apertures of three different radii (10 arcmin, 30 arcmin, and 10 comoving Mpc (cMpc)), to identify superclusters and voids.
The grid size of the density map was $1.5\times1.5$ arcmin$^2$, which was adopted for computing angular cross-correlations (\S\ref{s3}).
This study used density measurements based on an aperture size of $r=10$ cMpc.
We derived photometric redshifts and stellar masses of the targets based on the {\tt Mizuki} SED fitting code \citep{Tanaka2015,Tanaka2018}, where we set a chi-square limit of $\chi_\nu<5$, following suggestions by \citet[\S7]{Tanaka2018}.
Although the original density map was prepared for the HSC-SSP PDR2 over 360 deg$^2$, our study employed the updated catalogue for the HSC-SSP PDR3, which extended the survey area to 690 deg$^2$.
Consequently, 14.4 million objects were used for the density estimation.
Moreover, we adopted the CAMIRA red-sequence cluster catalogue \citep{Oguri2018} for further validations of the obtained density map.
\citet{Oguri2018} estimated the 3D richness ($N_\mathrm{mem}$) of red-sequence galaxies with M$_\star>10^{10.2}$ M$_\odot$ within $1h^{-1}$ proper Mpc using the cluster-finding algorithm, CAMIRA \citep{Oguri2014}.
We selected the CAMIRA cluster samples above a richness of $N_\mathrm{mem}=15$, approximately corresponding to a cluster virial mass $>10^{14}$ $h^{-1}$M$_\odot$ \citep{Oguri2018}.

\begin{figure}
\centering
\includegraphics[width=7.5cm]{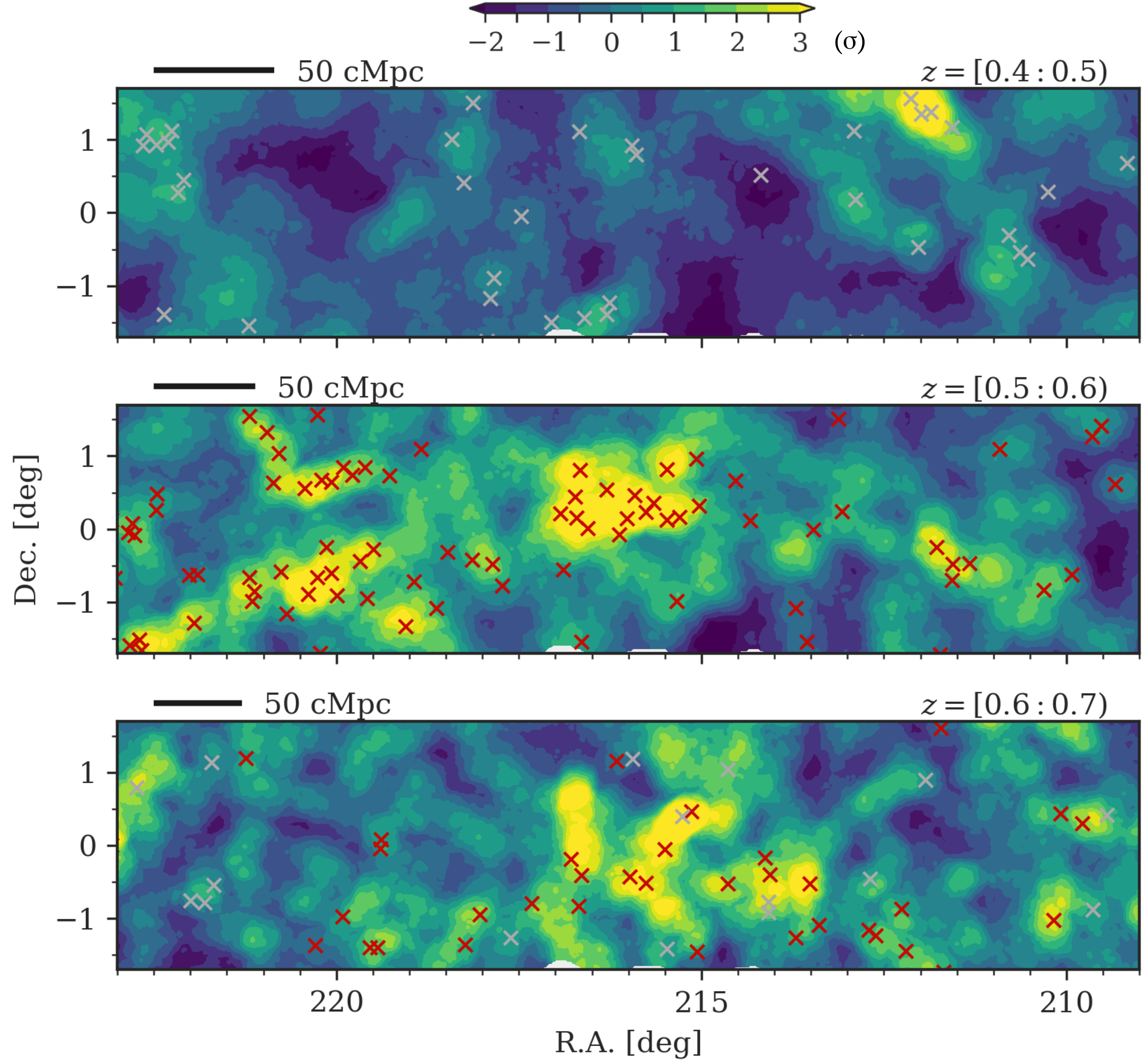}
\caption{
Projected 2D density map of the KGSc, from top to bottom, at $z=[0.4,0.5)$, $[0.5,0.6)$, and $[0.6,0.7)$.
Colourmaps depict overdensities $\sigma=[-2:+3]$ within $r=10$ cMpc apertures \citep{Shimakawa2021a}.
The cross symbols depict the CAMIRA red-sequence clusters \citep{Oguri2018}, with those at $z=[0.5,0.64)$ (see figure~\ref{fig2}) are highlighted in red.
}
\label{fig1}
\end{figure}

This study focused on the densest region at $z=[0.5:0.6)$ in the projected density map.
The three reasons behind our choice of this redshift slice for $z=$ 0.3--1.0 are as follows.
(1) The survey area can adequately capture large-scale structures on the $\sim100$ cMpc scale.
(2) We can reasonably estimate the stellar masses of galaxies \citep{Shimakawa2021b}.
(3) The approach is expected to detect WL signals with acceptable signal-to-noise ratios (SNRs), as reported by \citet{Shimakawa2021a}.
Following this, we searched for the most massive overdensity in the density map by measuring the surface area beyond a density excess of $\sigma=3$ (Area$_{\sigma>3}$).
Here, we applied a density-based clustering algorithm, DBSCAN \citep{Ester1996}, from a Python-based machine learning library, {\tt scikit-learn} (version 0.24.2; \citealt{Pedregosa2012}), to automate the selection process of overdense regions.

Consequently, we identified the largest supercluster at RA=$216.2^\circ$ and Dec=$-0.34^\circ$ with an Area$_{\sigma>3}=1321$ cMpc$^2$, referred to as the King Ghidorah Supercluster (KGSc), which is more than three times larger than any other overdense regions at the same redshift.
Figure~\ref{fig1} depicts the projected density map around the KGSc in three redshift slices from $z=0.4$ to $z=0.7$.
Surprisingly, the KGSc appears to be associated with 15 red-sequence clusters and multiple large-scale structures outside it.
Two large filamentous structures are present toward the northeast and southeast, and each system involves more than ten red-sequence clusters in the same redshift slice.
While additional overdensity peaks can be observed backwards along the line of sight, large-scale underdensities appear to spread in the foreground.

\begin{figure}
\centering
\includegraphics[width=7.5cm]{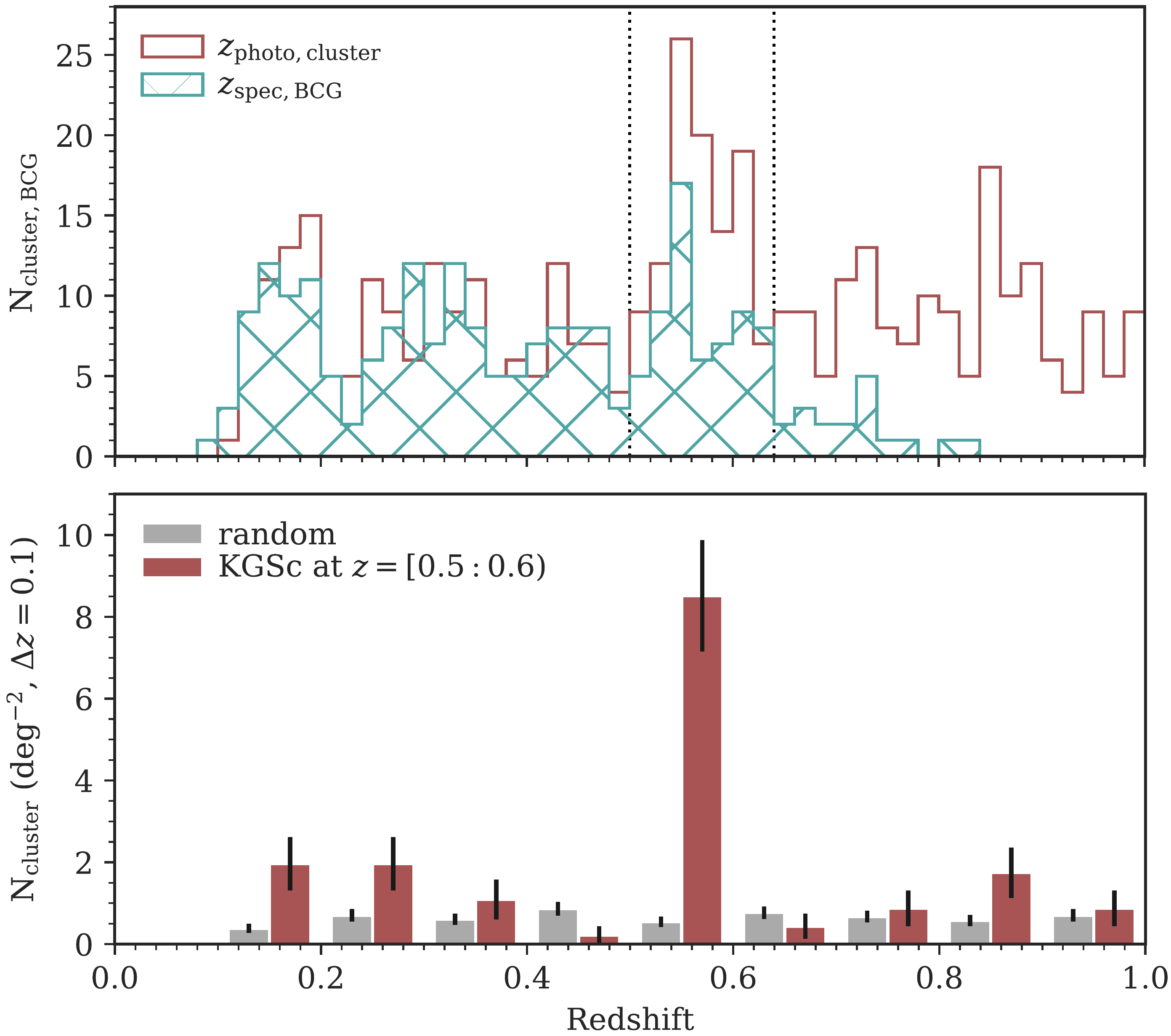}
\caption{
(Top) Distributions of photometric redshifts of the CAMIRA red-sequence clusters (shown in red) and the spectroscopic redshifts of the BCGs therein (shown in cyan hatch), as reported by \citet{Oguri2018}.
(Bottom) Number densities of the CAMIRA clusters in the KGSc ($\sigma>2$ in figure~\ref{fig1}, shown as red bars) and 50,000 random fields (grey bars).
The black vertical lines indicate Poisson noises in each redshift bin.
}
\label{fig2}
\end{figure}

We estimated the number densities of red-sequence clusters in the KGSc and neighbouring overdense structures compared to those in general fields.
The upper panel of figure~\ref{fig2} illustrates the redshift distributions of the CAMIRA clusters and the brightest cluster galaxies (BCGs) within figure~\ref{fig1}, which suggests that the entire system of the KGSc may spread out to $z\sim0.64$ (see also figure~\ref{fig1}).
Following this, we obtained the number densities of red-sequence clusters located in overdense areas above $\sigma=2$ within figure~\ref{fig1} and compared them with those within 50,000 random points from the entire survey field, except the KGSc region.
The obtained number densities demonstrated a notable excess at $z=[0.5:0.6)$ in the KGSc and approached $\delta=14.7\pm4.5$ compared to the mean density in random fields (figure~\ref{fig2}).
We also observed underdensities of the red-sequence clusters ($\delta=-0.75\pm0.26$) in the foreground at $z=[0.4:0.5)$.
We note that this number excess seen in low redshifts would be partially attributed to the Abell~1882 supergroup (RA=$213.6^\circ$ and Dec=$-0.4^\circ$ at $z=0.14$, \citealt{Abell1989}).

\section{Stellar mass and dark matter maps}
\label{s3}

In this section, we map and compare the stellar mass and dark matter distributions in and around the KGSc.
Notably, mapping the mass densities based on two independent measurements may help us validate massive structures and investigate the spatial correlations between the galaxies and dark matter in the supercluster.
The latter can be examined by analysing the WL signals and the colloquially named E-modes and B-modes, which are decomposed from the shear field \citep{Kaiser1992}. 
In particular, the E-mode map is treated as a gravitational overdensity map in WL studies.

\begin{figure}
\centering
\includegraphics[width=7.5cm]{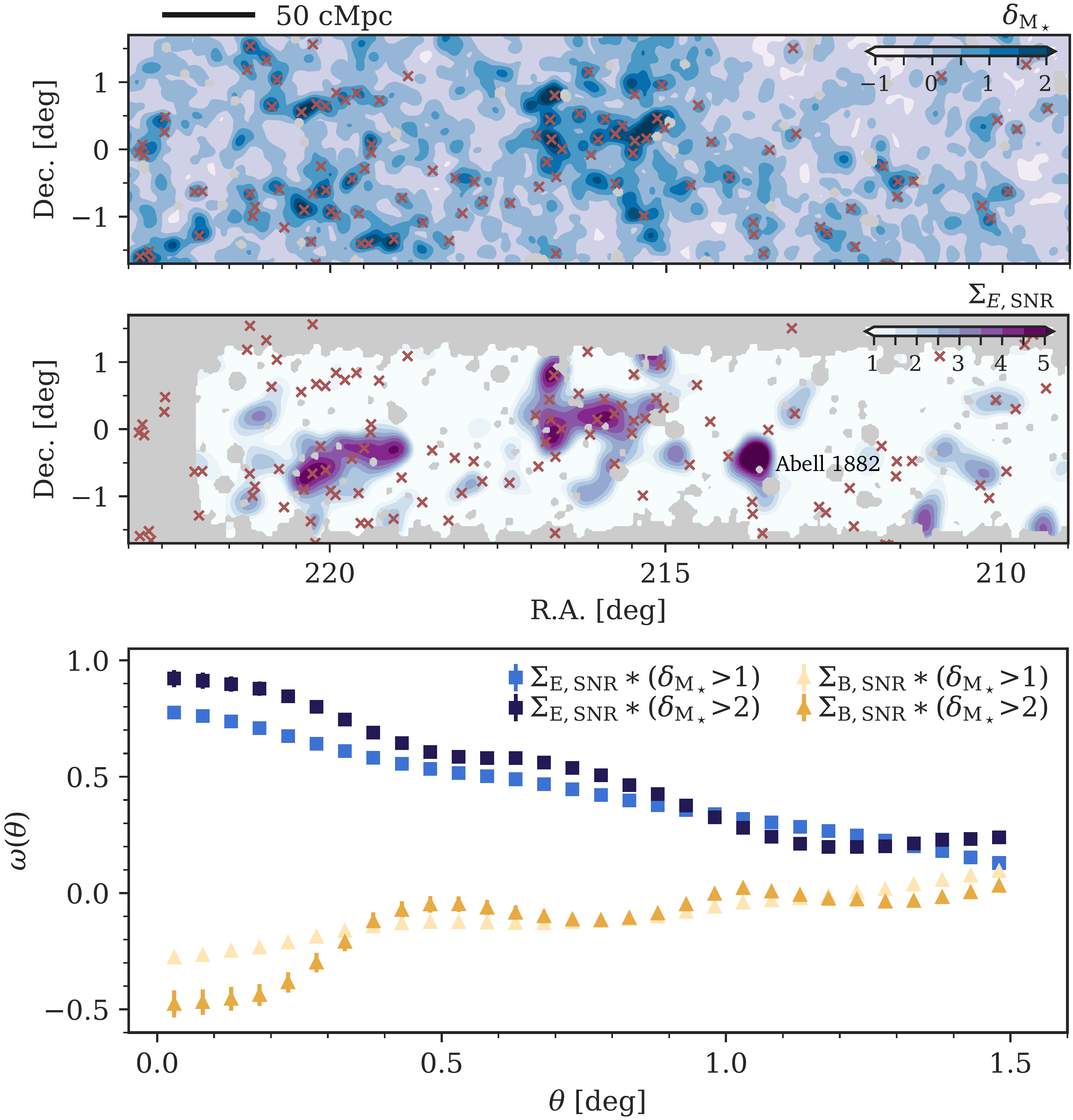}
\caption{
Projected 2D overdensities in the KGSc, traced by the (top) total stellar masses of massive galaxies (M$_\star>2.5\times10^{10}$ M$_\odot$), and (middle) SNR map of WL signals.
In both cases, we measured the overdensities using the Gaussian kernel with a beam width of 10 cMpc.
The cross symbols depict the red-sequence clusters at $z=[0.5,0.64)$ \citep{Oguri2018}.
The WL signal at RA=213.6$^\circ$ and Dec=$-$0.4$^\circ$ is attributed to the Abell~1882 supergroup at $z=0.14$.
(bottom) Angular cross-correlations between $\Sigma_\mathrm{E,SNR}$ and $\delta_\mathrm{M_\star}>1$ (light blue squares), $\Sigma_\mathrm{E,SNR}$ and $\delta_\mathrm{M_\star}>2$ (dark blue squares), $\Sigma_\mathrm{B,SNR}$ and $\delta_\mathrm{M_\star}>1$ (light yellow triangles), and $\Sigma_\mathrm{B,SNR}$ and $\delta_\mathrm{M_\star}>2$ (dark yellow triangles).
Refer to the text for a description of the calculation procedure.
}
\label{fig3}
\end{figure}

The WL analysis follows the analysis conducted by \citet{Okabe2019,Okabe2021}, who employed galaxy shape measurements based on the re-Gaussianization method \citep{Hirata2003}, which was implemented in the HSC-SSP pipeline.
The basis of such shape measurements is a Gaussian fitting with elliptical isophotes, which considers the profiles of both the point-spread function and the galaxy surface brightness (see \citealt{Rowe2015,Mandelbaum2018} for details).
Herein, we reconstructed a dimensional mass map $\Sigma(\theta)\simeq\kappa(\theta){\langle}\Sigma_\mathrm{cr}{\rangle}$ using a smoothing scale with a beam size of $=10$ cMpc following \citep{Okabe2008}.
This reconstruction was performed with calculation modifications to interpret the dimensional unit of $\Sigma_{\rm cr}$ based on its dimensionless counterpart \citep{Okabe2019,Okabe2021}.
Here, $\kappa$ denotes a dimensionless surface mass density, and $\langle \Sigma_{\rm cr}\rangle$ denotes an average critical surface mass density for background galaxies.
We assume a mean lens redshift $\langle z\rangle=0.55$ to suit the KGSc region. 
The background galaxies are selected based on $p=\int_{\langle z \rangle+0.2}^\infty p(z) dz >0.98$ \citep{Medezinski2018}, where $p(z)$ denotes a full probability function. 
The $\Sigma_{{\rm cr}}^{-1}$ values for individual background galaxies are then computed using $\int^\infty_{\langle z\rangle} \Sigma_{\rm cr}^{-1}(z)p(z)dz/\int^\infty_0 p(z)dz$.
The noise maps are computed based on 500 realisations of the random orientations of galaxy ellipticities with fixed positions.
The resulting map of the SNR ($\Sigma_\mathrm{E,SNR}$) is depicted in figure~\ref{fig3}.
Note that the SNR of the imaginary component of the weak-mass reconstruction is referred to as $\Sigma_\mathrm{B,SNR}$.

Subsequently, we calculated the total stellar mass $\Sigma_\mathrm{M_\star}$ and its variation $\delta_\mathrm{M_\star}$ for massive galaxies with M$_\star>2.5\times10^{10}$ M$_\odot$ in the KGSc based on the outputs obtained from the {\tt Mizuki} SED fitting (\citealt{Tanaka2018}, see also \S\ref{s2}).
We set the mass threshold to the original $i$-band magnitude limited samples ($i<23$) in consideration of the increasing systematic error at lower masses \citep{Shimakawa2021b}.
This decreased the sample size to 0.9 million in the entire field. 
Here, we emphasise that we used the same Gaussian kernel with a beam size of $10$ cMpc for the WL map to ensure a fair comparison between the stellar mass and dark matter distributions.

The obtained WL and total stellar mass maps present triple density peaks associated with the KGSc and also other peaks in the massive filament along the southeast (figure~\ref{fig3}).
While each of the two WL peaks along the south, among the three peaks associated with the main structure, involves three or four red-sequence clusters, the northeast peak has only one cluster with a richness of $N_\mathrm{mem}=23$, implying that blue galaxies and clusters may be missing.
In the entire filed, the lack of WL signals in the northeast overdensities suggests that this structure is less massive (or less concentrated) than the two others.
We note a WL peak at RA=213.6$^\circ$ and Dec=$-$0.4$^\circ$ owing to the Abell~1882 supergroup at $z=0.14$ \citep{Abell1989}.

We then obtained the pixel-to-pixel angular cross-correlations $\omega(\theta)$ between the stellar mass overdensity map ($\delta_\mathrm{M_\star}$) and the WL E-mode and B-mode maps (figure~\ref{fig3}) using the following formula: 
\begin{equation*}
\omega(\theta)=\frac{\Sigma_{i,j}\delta_i\delta_j}{\sqrt{\Sigma_{i^\prime}\delta_{i^\prime}^2\Sigma_{j^\prime}\delta_{j^\prime}^2}}
\quad \textrm{and} \quad
\delta\omega(\theta)^2 = \frac{1}{N^2}\sum_{k=1}^N(\omega(\theta)-\overline{\omega}(\theta))^2,
\end{equation*}
where $\delta\omega(\theta)$ denotes a measurement error, and $N$ denotes the number of pixel pairs ($i,j$) in each $\theta$ bin.
We assigned $\delta_\mathrm{M_\star}$ to $\delta_i$ and the variations in the SNRs of the E-mode ($\Sigma_\mathrm{E,SNR}$) and B-mode ($\Sigma_\mathrm{B,SNR}$) maps to $\delta_j$.
The derived $\omega(\theta)$ values are plotted in figure~\ref{fig3}, and these indicate a strong agreement between the overdensities of the stellar mass and dark matter.
These strong correlations are further corroborated by the decorrelations between the B-mode map and stellar mass overdensities.


\section{Discussion and conclusions}
\label{s4}

In this section, we present an analysis of the total masses of associated haloes using the mock catalogue dedicated to the HSC-SSP based on the ray-tracing cosmological simulation \citep{Takahashi2017,Shirasaki2017}.
Here, we should note that the mock catalogue covers the HSC-SSP PDR2 field over 430 deg$^2$, as it was originally established for our previous study \citep{Shimakawa2021a}.
Nevertheless, we consider that the survey area is sufficient for predicting the total mass of the KGSc, as described below.
The mock catalogue is formed by 14 simulation boxes, each comprising 2048$^3$ particles with sizes of 450 $h^{-1}$Mpc, based on the $N$-body code, {\tt GADGET2} \citep{Springel2005}.
Each simulation contains six independent realisations based on the initial power spectrum using the Code for Anisotropies in the Microwave Background ({\tt CAMB}; \citealt{Lewis2000}).
This enables a large-volume simulation suitable for extremely wide-field surveys such as the HSC-SSP.
Note that the corresponding dark matter halo catalogue was created by the halo identifier, {\tt ROCKSTAR} \citep{Behroozi2013}.
The simulation satisfactorily resolves dark matter haloes of the virial mass down to $\sim2\times10^{12}$ M$_\odot$ at $z=$ 0.5--0.6.
Additional details pertaining to the simulation and the mock catalogue can be found in \citet{Takahashi2017} and \citet{Shimakawa2021a}, respectively (see also Appendix~\ref{a1}).

\begin{figure}
\centering
\includegraphics[width=7.5cm]{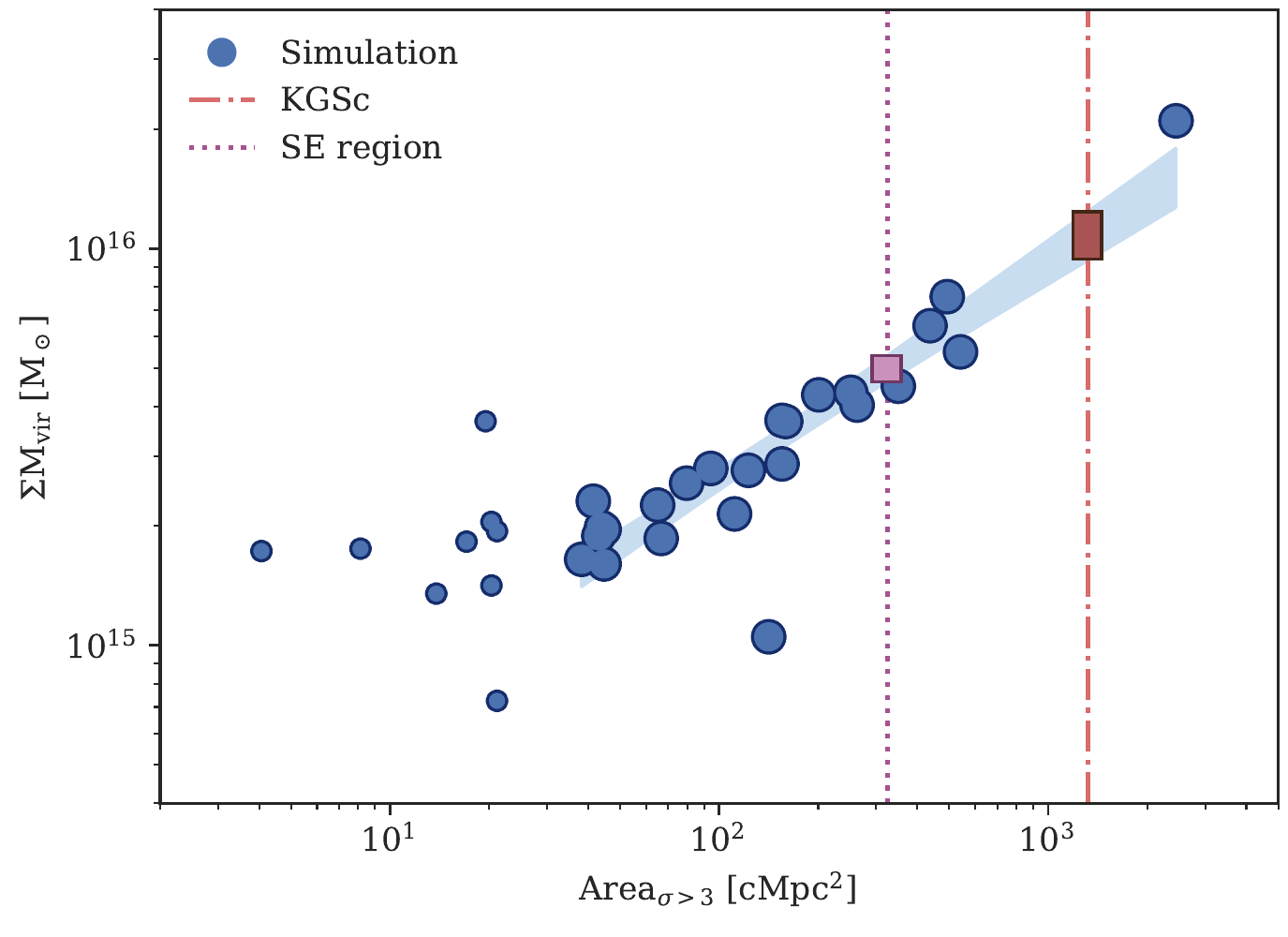}
\caption{
Blue circles depict the total masses ($\Sigma\mathrm{M_{vir}}$) of massive haloes ($\gtrsim2\times10^{12}$ M$_\odot$) associated with overdensities ($\sigma>3$) against areas obtained from the mock catalogue.
The blue filled region depicts the $\pm1\sigma$ area around the best-fit line for the mock data with log overdensity areas $>1.5$ (larger circles).
The red dotted line indicates the overdensity area of the KGSc, which suggests that the KGSc may have a total mass of $(1.09\pm0.15)\times10^{16}$ M$_\odot$.
}
\label{fig4}
\end{figure}

As our previous study \citep{Shimakawa2021a} established the projected 2D density map within $r=10$ cMpc apertures based on the mock catalogue in a manner equivalent to an observation, the present study focuses on measuring the overdensity areas (Area$_{\sigma>3}$) for the mock overdensity map at $z=[0.5:0.6)$, as described in \S\ref{s2}.
We then obtain the total mass of dark matter haloes associated with selected overdensities within the range of $r=10$ cMpc in the same redshift space.
Consequently, the derived total masses are found to be tightly correlated with the areas at Area$_{\sigma>3}\gtrsim30$ cMpc$^2$ (figure~\ref{fig4}).
The best-fit line considered by the curve fitting code, {\tt lmfit} \citep{Newville2021}, suggests that the KGSc and the southeast (SE) region may attain the total masses of $(1.09\pm0.15)\times10^{16}$ M$_\odot$ and $(5.00\pm0.38)\times10^{15}$ M$_\odot$, respectively.

In summary, we discovered a supercluster, termed as the KGSc, at $z=0.55$.
This supercluster not only hosts 15 red-sequence clusters within $\Delta z=0.1$ but also involves multiple other structures, including two massive filaments across a 400 cMpc scale.
The presence of the KGSc is also confirmed based on the total stellar mass of associated galaxies and the WL mass maps.
These two independently derived mass distributions are found to be tightly coupled.
The large-volume simulation suggests that the KGSc is one of the most massive structures in the universe, accounting for $1\times10^{16}$ M$_\odot$ of the total mass of the associated dark matter haloes already present at $z=0.55$.
Further, precise measurements of halo masses of the member clusters and cross-correlations between galaxies and dark matter require a wide-field multi-object spectrograph such as the Subaru Prime Focus Spectrograph \citep{Takada2014}.
However, understanding the entire picture even with spectroscopic data still remains a challenge to overcome, given the difficulty in resolving the fingers-of-god effect.
A novel solution is thus required to scrutinise the mass assembly histories around superclusters.


\section*{Acknowledgements}

This research is based on data collected at Subaru Telescope, which is operated by the National Astronomical Observatory of Japan.
We are honoured and grateful for the opportunity of observing the Universe from Maunakea, which has the cultural, historical and natural significance in Hawaii.

The Hyper Suprime-Cam (HSC) collaboration includes the astronomical communities of Japan and Taiwan, and Princeton University. The HSC instrumentation and software were developed by the National Astronomical Observatory of Japan (NAOJ), the Kavli Institute for the Physics and Mathematics of the Universe (Kavli IPMU), the University of Tokyo, the High Energy Accelerator Research Organization (KEK), the Academia Sinica Institute for Astronomy and Astrophysics in Taiwan (ASIAA), and Princeton University. Funding was contributed by the FIRST program from Japanese Cabinet Office, the Ministry of Education, Culture, Sports, Science and Technology (MEXT), the Japan Society for the Promotion of Science (JSPS), Japan Science and Technology Agency (JST), the Toray Science Foundation, NAOJ, Kavli IPMU, KEK, ASIAA, and Princeton University. 
This paper makes use of software developed for the Large Synoptic Survey Telescope. We thank the LSST Project for making their code available as free software at \url{http://dm.lsst.org}.
The Pan-STARRS1 Surveys (PS1) have been made possible through contributions of the Institute for Astronomy, the University of Hawaii, the Pan-STARRS Project Office, the Max-Planck Society and its participating institutes, the Max Planck Institute for Astronomy, Heidelberg and the Max Planck Institute for Extraterrestrial Physics, Garching, The Johns Hopkins University, Durham University, the University of Edinburgh, Queen’s University Belfast, the Harvard-Smithsonian Center for Astrophysics, the Las Cumbres Observatory Global Telescope Network Incorporated, the National Central University of Taiwan, the Space Telescope Science Institute, the National Aeronautics and Space Administration under Grant No. NNX08AR22G issued through the Planetary Science Division of the NASA Science Mission Directorate, the National Science Foundation under Grant No. AST-1238877, the University of Maryland, and Eotvos Lorand University (ELTE) and the Los Alamos National Laboratory.

We thank anonymous referee for helpful feedback.
We would like to thank Editage (\url{www.editage.com}) for English language editing.
This work is supported by MEXT/JSPS KAKENHI Grant Numbers (19K14767, 20H05861).
Numerical computations were in part carried out on Cray XC50 at Center for Computational Astrophysics, National Astronomical Observatory of Japan.
This work made extensive use of the following tools, {\tt NumPy} \citep{Harris2020}, {\tt Matplotlib} \citep{Hunter2007}, {\tt TOPCAT} \citep{Taylor2005}, {\tt Astopy} \citep{AstropyCollaboration2013}, and {\tt pandas} \citep{Reback2021}.

\section*{Data Availability}

The catalogue and data underlying this article are available on the public data release site of Hyper Suprime-Cam Subaru Strategic Program (\url{https://hsc.mtk.nao.ac.jp/ssp/data-release/}).



\bibliographystyle{mnras}
\bibliography{rs22b} 



\appendix

\section{Superclusters in the mock data}\label{a1}

This appendix provides a supplementary explanation on the most massive structures in the mock data (figure~\ref{fig4}).
For further details on the cosmological simulation and the mock catalogue, readers may refer to \citet{Takahashi2017,Shimakawa2021a}.

\begin{figure}
\centering
\includegraphics[width=7.5cm]{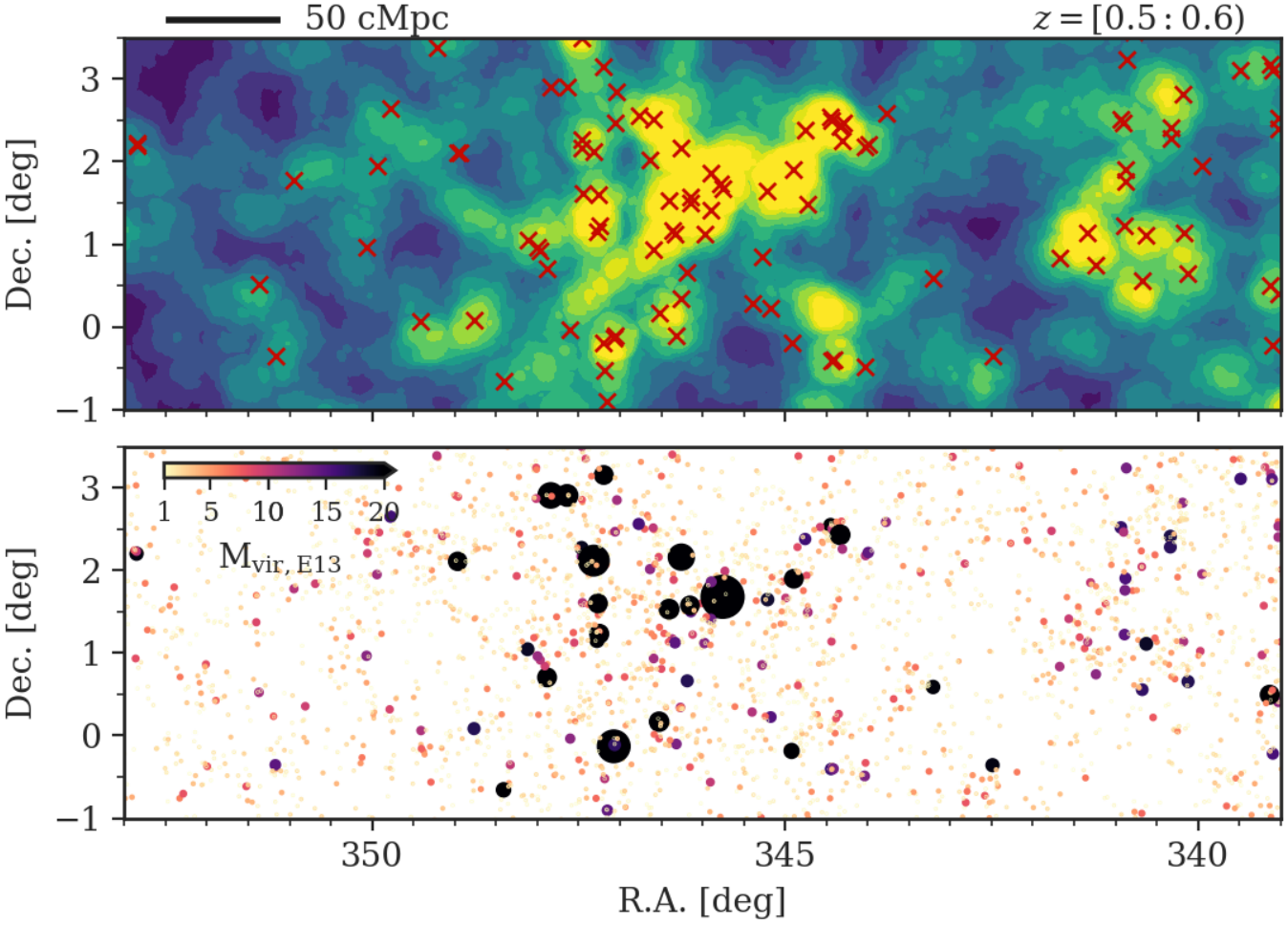}
\caption{
(Top) Same as figure~\ref{fig1} but for the most massive supercluster in the mock data at $z=[0.5,0.6)$.
The red cross symbols indicate the positions of cluster haloes with M$_\mathrm{vir}\geq1\times10^{14}$ M$_\odot$.
(Bottom) Projected 2D distribution of massive dark matter haloes with M$_\mathrm{vir}\geq1\times10^{13}$ M$_\odot$ (and up to $7\times10^{14}$ M$_\odot$) in the same redshift slice.
The marker sizes and colours are scaled by their halo masses as indicated in the inset colour bar.
}
\label{fig1a}
\end{figure}

We determined that the abundance of supercluster candidates with an Area$_{\sigma>3}>100$ cMpc$^2$ in the mock sample is $(4.0\pm1.0)\times10^{-2}$ per deg$^2$ (figure~\ref{fig4}), which is consistent with that in HSC-SSP PDR3 ($(4.8\pm0.8)\times10^{-2}$ per deg$^2$). 
This implies that a field coverage of more than 20--30 deg$^2$ is generally required to locate at least one of these massive structures in this redshift space based on a blind survey.
However, we note that the uncertainty in the number density may be underestimated owing to the lack of independent $N$-body realisations, as it only provides a sense of typical statistical fluctuations.
For more precise error estimates, more $N$-body simulations may be required in future research.
Figure~\ref{fig1a} presents a projected 2D density map around the most massive structure at $z=[0.5:0.6)$ with an Area$_{\sigma>3}=2438$ cMpc$^2$ in the mock data.
The figure also presents massive haloes in the corresponding region, resembling the projected density map and the cluster distribution obtained from the actual observation based on the HSC-SSP PDR3 (figure~\ref{fig1}).
This mock supercluster is more massive ($\gtrsim\times3$) than the others in the cosmological simulation over 430 deg$^2$. 
Given that only one similar-sized supercluster was discovered in the HSC-SSP PDR3 over 690 deg$^2$, such a gigantic structure with an Area$_{\sigma>3}>1000$ cMpc$^2$ and $\Sigma\mathrm{M_{vir}}\ge1\times10^{16}$ M$_\odot$ may be extremely rare ($\sim2.0\times10^{-3}$ per deg$^2$ or less).

\bsp	
\label{lastpage}
\end{document}